\documentclass[a4paper,oneside]{amsart}

\newcommand{\varkappa}{\kappa}

\renewcommand{\det}{{\rm det}}

\newcommand{\mC}{\mathbb{C}}

\newcommand{\cF}{\mathcal{F}}

\newcommand{\cH}{\mathcal{H}}

\newcommand{\mN}{\mathbb{N}}

\newcommand{\bp}{\mathbf{p}}

\newcommand{\bq}{\mathbf{q}}

\newcommand{\mR}{\mathbb{R}}

\newcommand{\bx}{\mathbf{x}}

\DeclareMathOperator{\spec}{spec}

\newtheorem{thm}{Theorem}
\newtheorem{cor}[thm]{Corollary}

\theoremstyle{remark}

\begin{document}

\title[]{On the number of bound states
for weak perturbations of spin-orbit Hamiltonians}

\author{Jochen Br\"uning}
\address{Institut f\"ur Mathematik, Humboldt-Universit\"at zu Berlin,
Rudower Chaussee 25, 12489 Berlin, Germany}
\email{bruening@mathematik.hu-berlin.de}

\author{Vladimir Geyler}
\address{Institut f\"ur Mathematik, Humboldt-Universit\"at zu Berlin,
Rudower Chaussee 25, 12489 Berlin, Germany \&
Mathematical Faculty, Mordovian State University, 430000 Saransk, Russia}
\email{geyler@mathematik.hu-berlin.de}

\author{Konstantin Pankrashkin}

\address{Institut f\"ur Mathematik, Humboldt-Universit\"at zu Berlin,
Rudower Chaussee 25, 12489 Berlin, Germany \&
D\'epartement de Math\'ematiques, Universit\'e Paris 13,
99 avenue Jean-Baptiste Cl\'ement, 93430 Villetaneuse, France}
\email{const@math.univ-paris13.fr}

\begin{abstract}
We give a variational proof of the existence of infinitely many
bound states below the continuous spectrum for some weak
perturbations of a class of spin-orbit Hamiltonians including the
Rashba and Dresselhaus Hamiltonians.
\end{abstract}

\maketitle

In the recent paper \cite{CM} Chaplik and Magarill have discovered a
surprising fact: the Rashba Hamiltonian $H_R$,
\[
H_R=\begin{pmatrix}
p^2 & \alpha_R(p_y+ip_x)\\
\alpha_R(p_y-ip_x) & p^2
\end{pmatrix}
\]
($\alpha_R$ is the Rashba constant expressing the strength of the
spin-orbit coupling \cite{BR,Win}) perturbed by a
short-range rotationally symmetric negative potential has an
infinite number of eigenvalues below the threshold of the
continuous spectrum. More precisely, for a rotationally symmetric
shallow potential well $V$ ($mUR^2/\hbar^2\ll 1$, where $m$ is the
effective mass and $U$ and $R$ are the depth and radius of the
well, respectively), in~\cite{CM} a system of equations was
derived to be satisfied by the eigenvalues,
and it was shown (using the pole approximation for the calculation of
some integrals) that the system has an infinite number of
solutions below the continuous spectrum of $H_R+V$.

In the present note we are going to provide a strict mathematical
justification for the existence of infinitely many bound states
below the continuous spectrum for short-range perturbations of a
much larger class of spin-orbit Hamiltonians, which includes, in
particular, the above Rashba Hamiltonian as well as the
Dresselhaus Hamiltonian~\cite{Win},
\[
H_D=
\begin{pmatrix}
p^2 & -\alpha_D(p_x+i p_y)\\
-\alpha_D(p_x-ip_y) & p^2
\end{pmatrix}
\]
(here $\alpha_D$ is the Dresselhaus constant). Our technique is
elementary and uses the max-min principle in the spirit
of~\cite{YdL}. Moreover, the potential well $V$ can be non-symmetric,
and ``shallow'' in our context means $V\in L^1$.

Denote by $\cH$ the Hilbert space
$L^2(\mR^2)\otimes\mC^2$ of two-dimensional spinors; by $\cF$ we
denote the Fourier transform $\cF:\,L^2(\mR^2)\rightarrow
L^2(\mR^2)$; then $\cF_2:=\cF\otimes 1_{\mC^2}$ is the Fourier
transform in $\cH$. Let $H_0$ be the self-adjoint operator in
$\cH$ whose Fourier transform $\widehat H_0:=\cF_2H_0\cF_2^{-1}$ is the
multiplication by the matrix
\begin{equation}
                \label{S1}
\widehat H_0(\bp)=\begin{pmatrix} p^2& A(\bp)\\
                  A^*(\bp)& p^2\\
                  \end{pmatrix}\,,\quad\,\bp\in\mR^2\,,
\end{equation}
where $A$ is a continuous complex function on $\mR^2$,
star $*$ means the complex conjugation and, as usual, $p:=|\bp|$.
It is obvious that $H_0$ is self-adjoint. Clearly, the Rashba and Dresselhaus
Hamiltonians
have the form \eqref{S1} with a linear $A$. In generalizing the linearity we assume
\begin{equation}
\label{A1} \limsup\limits_{p\to \infty}\dfrac{|A(\bp)|}{p^2}<1
\end{equation}
Clearly, $H_0$ has no discrete spectrum; its spectrum is the union
of images of two functions $\lambda_{\pm}$ (dispersion laws):
$\lambda_\pm(\bp)=p^2\pm|A(\bp)|$, hence $\spec
H_0=[\varkappa,+\infty)$, where
$\varkappa:=\inf\{p^2-|A(\bp)|\,:\,\bp\in\mR^2\}>-\infty$.
Moreover, there is a unitary matrix $M(\bp)$ depending
continuously on $\bp\in\mR^2$ such that
\begin{equation}
                \label{S3}
M(\bp)\widehat H_0(\bp)M^*(\bp)=
                 \begin{pmatrix} \lambda_+(\bp)&0\\
                  0& \lambda_-(\bp)\\
                  \end{pmatrix},\quad\,\bp\in\mR^2\,.
\end{equation}

Denote $S:=\{\bp\in\mR^2:\,\lambda_-(\bp)=\varkappa\}$; this is a
non-empty compact set. We will assume that
\begin{equation}
\label{A2}
\text{the function $|A(\bp)|$ is of class $C^2$ in a
neighborhood of $S$.}
\end{equation}
For the Rashba and Dresselhaus Hamiltonians one has
$\varkappa=-\alpha_J^2/4$ ($J=R,D$) and $S$ is the circle
$\{\bp: 2p=|\alpha_J|\}$; in these cases $S$ is called {\it the
loop of extrema}. The condition (\ref{A2}) is obviously satisfied
for these Hamiltonians.

The two condtions \eqref{A1} and \eqref{A2} imply that for every
$\bp_0\in S$ there is a constant $c>0$ such that we have
\begin{equation}
                \label{S4}
0\le\lambda_-(\bp)-\varkappa\le c(\bp-\bp_0)^2\,
\end{equation}
for all $\bp\in\mR^2$.

Let now $V$ be a real-valued scalar potential from $L^p(\mR^2)$
with some $p>1$. Using the Sobolev inequality and an explicit form
for the Green function of $-\Delta$ we see that
$V(-\Delta+E)^{-1}$ with $E>0$ is a Hilbert--Schmidt operator;
therefore, $V$ is a compact perturbation of
$(-\Delta)\oplus(-\Delta)$ (we denote $V\oplus V=VI_{\mC^2}$ where
$I_{\mC^2}$ is the identity operator in $\mC^2$ by the symbol $V$
since this notation does not lead to confusion). Using \eqref{A1}
it is easy to show that the domains of $(-\Delta)\oplus(-\Delta)$
and $H_0$ coincide and the graph norms in these domains are
equivalent. Hence, $V$ is a compact perturbation of $H_0$. As a
result, we get that the operator $H:=H_0+V$ is well defined and
$\spec_\text{ess}H=[\varkappa, +\infty)$.

Below, for a distribution $f$ by $\widehat f$ we denote its Fourier transform.
To avoid mixing terminology, an Hermitian $n\times n$ matrix $C$ will be called
\emph{positive definite} if $\langle \xi|C\xi\rangle>0$ for any non-zero $\xi\in\mC^n$,
and will be called \emph{positive semi-definite} if the above equality is non-strict.
By analogy one introduces \emph{negative definite} and \emph{negative semi-definite}
matrices.

\begin{thm}\label{th1}
Let $N\in\mN$. Assume that $V\in L^1(\mR^2)$ and that $\widehat V$
satisfies the following condition: there are $N$ points
$\bp_1,\ldots\bp_N\in S$ such that the matrix $\big(\widehat
V(\bp_m-\bp_n)\big)_{1\le m,n\le N}$ is negative definite. Then
$H$ has at least $N$ eigenvalues, counting multiplicity, below
$\varkappa$.
\end{thm}

\begin{proof} According to the max-min
principle, it is sufficient to show that  we
can find $N$ vectors $\Psi_m\in\cH$, $m=1,\ldots,N$, such that the
matrix with the entries $\langle\Psi_m|(H-\varkappa)\Psi_n\rangle$, $1\le m,n\le N$.
is negative definite (the vectors $\Psi_m$ are then a posteriori
linearly independent).

Denote $\displaystyle
f_a(\bx):=\exp\left(-\dfrac{1}{2}|\bx|^a\right)$, $\bx\in\mR^2$,
with $a>0$. As observed in~\cite{YdL}
\begin{equation}
              \label{S5}
\int_{\mR^2}\big|\nabla\,f_a(\bx)\big|^2\,d\bx=\frac{\pi}{2}a.
\end{equation}
Furthemore, by the Lebesgue majorization theorem,
\begin{equation}
              \label{S6}
\lim\limits_{a\to0}\,\int_{\mR^2} V(\bx)\,|f_a(\bx)|^2\,d\bx=e^{-1}\,\int_{\mR^2}
V(\bx)\,d\bx\,.
\end{equation}

Let $\widehat f_a(\bp)$ be the Fourier transform of $f_a(\bx)$.
Take spinors $\Psi_m$ such that their Fourier transforms
$\widehat\Psi_m$ are of the form
$\widehat\Psi_m(\bp)=M(\bp)\psi_m(\bp)$, where
\begin{equation}
              \label{S7}
\psi_m(\bp)=\left(\begin{matrix}0\\
\widehat
f_a(\bp-\bp_m)\\\end{matrix}\right)
\end{equation}
and $M(\bp)$ is taken from \eqref{S3}.
We show that if $a$ is sufficiently small, then the
matrix $\big(\langle
\Psi_m|(H-\varkappa)\Psi_n\rangle\big)$ is
negative definite.
For this purpose it is sufficient to show that
\begin{gather}
\label{f1}
 \lim_{a\to0}\langle \Psi_m|(H_0-\varkappa)\Psi_n\rangle=0,\\
\label{f2} \lim_{a\to0}\langle \Psi_m|V\Psi_n\rangle=2\pi
e^{-1}\widehat V(\bp_m-\bp_n)
\end{gather}
for all $(m,n)$.

By definition of $\Psi_m$ one has
\begin{gather*}
\big|\langle
\Psi_m|(H_0-\varkappa)\Psi_n\rangle\big|=\Big|\int_{\mR^2}
(\lambda_-(\bp)-\varkappa){\widehat f^*_a(\bp-\bp_m)}\widehat
f_a(\bp-\bp_n)\,d\bp\Big|\\
\le\left[\int_{\mR^2} \big(\lambda_-(\bp)-\varkappa\big)
\,\big|\widehat f_a(\bp-\bp_m)\big|^2\,d\bp\right]^{\frac{1}{2}}
\left[\int_{\mR^2}
\big(\lambda_-(\bp)-\varkappa\big)\,\big|\widehat
f_a(\bp-\bp_n)\big|^2\,d\bp\right]^{\frac{1}{2}}.
\end{gather*}
On the other hand, by \eqref{S4} and \eqref{S5} one has
\begin{multline*}
0\le \int_{\mR^2} \big(\lambda_-(\bp)-\varkappa\big)\,\big|\widehat f_a(\bp-\bp_m)\big|^2\,d\bp\\
\le c\int_{\mR^2} (\bp-\bp_m)^2\,\big|\widehat f_a(\bp-\bp_m)\big|^2\,d\bp
= c\int_{\mR^2} \bp^2\big|\widehat f_a(\bp)\big|^2\,d\bp=\frac{\pi}{2}ca,
\end{multline*}
which proves \eqref{f1}.
As for \eqref{f2}, one has
\begin{multline*}
\langle \Psi_m|V\Psi_n\rangle= \int_{\mR^2}\int_{\mR^2}\big\langle\widehat
\Psi_m(\bp)\big|\widehat V(\bp-\bq)\widehat\Psi_n(\bq)\big\rangle\,d\bp\,d\bq\\
=\int_{\mR^2}\int_{\mR^2}\big\langle\psi_m(\bp)\big|\widehat V(\bp-\bq)\psi_n(\bq)\big\rangle\,d\bp\,d\bq
\end{multline*}
since matrices $\widehat V(\bp-\bq)$ and $M(\bp)$ commute. On the
other hand,
\begin{multline*}
\int_{\mR^2}\int_{\mR^2}\big\langle\psi_m(\bp)\big|\widehat V(\bp-\bq)
\psi_n(\bq)\big\rangle\,d\bp\,d\bq\\
=\int_{\mR^2}\int_{\mR^2} \widehat V(\bp-\bq){\widehat
f^*_a(\bp-\bp_m)}\widehat
f_a(\bq-\bp_n)\,d\bp\,d\bq\\
= \int_{\mR^2} V(\bx)e^{i(\bp_m-\bp_n)\bx}\big|f_a(\bx)\big|^2\,d\bx
\stackrel{a\to 0}{\longrightarrow} 2\pi
e^{-1}\widehat V(\bp_m-\bp_n).
\end{multline*}
The proof is complete.
\end{proof}

Let us list several corollaries.

\begin{cor}\label{cor}
If  $\displaystyle \int_{\mR^2}
V(\bx)\,d\bx<0$, then $H$ has at least one eigenvalue below
$\varkappa$.
\end{cor}

\begin{proof}
Since $S$ is non-empty, it remains to note that
$\widehat V(\mathbf{0})\equiv \displaystyle\int_{\mR^2} V(\bx)\,d\bx$.
\end{proof}
Note that taking $A=0$ we recover a result of
B.~Simon: A weak negative perturbation of the free Hamiltonian $-\Delta$
always has a bound state below the threshold of the continuous
spectrum \cite{Sim}.

Below by $\# S$ we denote the number of points in $S$, if $S$ is finite,
and $\infty$, otherwise.
\begin{cor}\label{boch} Let $V$ be non-positive and non-vanishing on a set of positive measure.
Then $H$ has at least $\# S$ eigenvalues below $\kappa$ counting multiplicities.
\end{cor}

\begin{proof} It is sufficient to show that the matrix $\big(\widehat V(\bp_m-\bp_n)\big)_{1\le m,n\le N}$ is
negative definite for every choice of points
$\bp_1,\ldots,\bp_N\in\mR^2$. By the Bochner theorem,
$\displaystyle-\sum_{mn}\widehat V(\bp_m-\bp_n)\xi_m^*\xi_n\ge0$
for any $(\xi_m)\in\mC^N$ and it remains to note that
$\displaystyle\sum_{mn}\widehat V(\bp_m-\bp_n)\xi_m^*\xi_n\ne0$
for $(\xi_m)\ne0$. In fact, if $\displaystyle\sum_{mn}\widehat
V(\bp_m-\bp_n)\bar\xi_m\xi_n=0$, then
\[
\int_{\mR^2}\Big|\sum_{m}\xi_me^{i\bp_m \bx}\Big|^2\,V(\bx)\,d\bx=0\,;
\]
therefore, $\displaystyle\sum_{m}\xi_me^{i\bp_m \bx}=0$ on the
support of $V$. Since exponents $e^{i\bp_m \bx}$ are real-analytic in $\bx$ and $\displaystyle\sum_{m}\xi_me^{i\bp_m \bx}=0$ on a
set of non-zero Lebesgue measure, the equality
$\displaystyle\sum_{m}\xi_me^{i\bp_m \bx}=0$ is valid everywhere
on $\mR^2$. On the other hand, $e^{i\bp_m \bx}$ are linearly
independent, and we obtain $\xi_m=0$ for all $m$.
\end{proof}

By Corollary~\ref{boch}, perturbations of both the Rashba and
Dresselhaus Hamiltonians by negative potentials from $L^p\cap
L^1$, $p>1$, have infinitely many eigenvalues below the threshold of the
continuous spectrum. Another important example where
Corollary~\ref{boch} can be applied is the Hamiltonian with both
Rashba and Dresselhaus terms:
\[
H_{RD}=\begin{pmatrix}
p^2 & \alpha_R(p_y+ip_x)-\alpha_D(p_x+i p_y)\\
\alpha_R(p_y-ip_x)-\alpha_D(p_x-ip_y) & p^2
\end{pmatrix},
\]
which is used for describing the ballistic spin transport through a
two-dimensional mesoscopic metal/semiconductor/metal double
junction in the presence of spin-orbit interaction \cite{LC}. In
this case $\kappa=(|\alpha_R|^2+|\alpha_D|^2)/4$ and $S$ contains
exactly two points: $S=\{\bp_0,-\bp_0\}$, where
\[
\bp_0=\begin{cases}
               \displaystyle\frac{1}{2\sqrt{2}}(\alpha_R+\alpha_D,-\alpha_R-\alpha_D)
               \,,&\quad\text{if}\quad \alpha_R\alpha_D>0,\\
               \noalign{\medskip}
               \displaystyle\frac{1}{2\sqrt{2}}(\alpha_R-\alpha_D,\alpha_R-\alpha_D)
               \,,&\quad\text{if}\quad \alpha_R\alpha_D<0\,.\\
\end{cases}
\]
In virtue of Theorem~\ref{th1}, in this case $H_{RD}+V$ has at
least two eigenvalues below $\varkappa$.

We note that Theorem~\ref{th1} delivers some quantitative
information on the eigenvalues. If
$\mu_1\le\mu_2\le\ldots\le\mu_N$ are the solutions to
$\det\big(\textsf{V}_a-\mu \textsf{G}_a\big)=0$, where
$\textsf{V}_a$ (respectively, $\textsf{G}_a$) is the matrix with
the entries $\langle\Psi_m|V\Psi_n\rangle$ (respectively,
$\langle\Psi_m|\Psi_n\rangle$), $1\le m,n\le N$, then, for
sufficient small $a$, the $n$-th eigenvalue $E_n$ of $H$ ($1\le n
\le N$) obeys the estimate $E_n\le \mu_n<\kappa$.

It is worth noting that the class of perturbations for which the
above machinery works contains singular perturbations supported on
sets of zero Lebesgue measure \cite{BEKS}, in particular, the
Dirac $\delta$-functions supported by curves. The last class of
``potentials'' was used, e.g. in \cite{CM1} for studying the
effect of spin-orbit interaction on bound states of electrons.
Nevertheless, accurate demonstrations in this case require rather
cumbersome purely technical details and are outside of the scope
of the note. We remark only that $H_R$ or $H_D$ perturbed by the
Dirac $\delta$-function supported by a circle has infinite number
of eigenvalues below the threshold of the continuum spectrum. On
the other hand, point perturbations of these Hamiltonians with
one-point supports have exactly one bound state below the
continuum.

\subsection*{Acknowledgments}
We are very grateful to A.~V.~Chaplik and L.~A.~Chernozatonskii
for discussion and valuable remarks. J.B. and V.G. were supported
in part by the Sonderfoschungsbereich 647 ``Space, Time, Matter''
and the program of German-Russian cooperation of the Deutsche
Forschungsgemeinschaft (Kz. 436 RUS 113/785). K.P. was supported
by the research fellowship of the Deutsche Forschungsgemeinschaft (PA
1555/0-1).


\begin{thebibliography}{99}

\bibitem{BEKS} J.~F.~Brasche, P.~Exner, Yu.~A.~Kuperin, P.~\v
        Seba: Schr\"odinger operators with singular interactions.
        \emph{J. Math. Anal. Appl.} {\bf184}, 112--139 (1994).


\bibitem{BR} Yu.~A.~Bychkov, E.~I.~Rashba:
Properties of a 2D electron gas with lifted spectral degeneracy,
\emph{ JETP Lett.} {\bf39}, 78--81 (1984).

\bibitem{CM} A.~V.~Chaplik, L.~I.~Magarill:
Bound states in a two-dimensional short range potential induced by
spin-orbit interaction, \emph{Phys. Rev. Lett.} {\bf96}, 126402
(2006).

\bibitem{CM1} A.~V.~Chaplik, L.~I.~Magarill:
Spin-dependent localization of electrons in low-dimensional
systems, \emph{Physica E} {\bf34}, 344--347 (2006).


\bibitem{LC} M.~Lee, M.-S.~Choi: Ballistic spin currents in
mesoscopic metal/In(Ga)As/metal junctions, \emph{Phys. Rev. B.}
{\bf71}, 153306 (2005).


\bibitem{Sim} B.~Simon: The bound state of weakly coupled
Schr\"odimger operators in one and two dimensions, \emph{Ann.
Phys.} {\bf 97}, 279--288 (1976).

\bibitem{Win} R.~Winkler: {\it Spin-orbit coupling effects in
two-dimensional electron and hole systems}, Springer, Berlin etc.,
2003.

\bibitem{YdL} K.~Yang, M.~de~Llano:
Simple variational proof that any two-dimensional potential well
supports at least one bound state, \emph{Am. J. Phys.} {\bf 57},
85--86 (1989).

\end{thebibliography}
\end{document}